\documentclass[aps,prd,superscriptaddress,showpacs,preprint,amsmath,amssymb]{revtex4}
\usepackage{graphicx, bm}
\usepackage[usenames]{color}
\usepackage{slashed}

\usepackage{subcaption}
\usepackage{slashed}
\usepackage{float}
\usepackage{array}
\begin{document}

\draft

\title{Probing CP-violating Higgs-gauge Boson Couplings at Future Muon Collider}

\author{Emre Gurkanli\footnote{egurkanli@sinop.edu.tr}}
\affiliation{\small Department of Physics, Sinop University, Türkiye.\\}

\author{Serdar Spor\footnote{serdar.spor@beun.edu.tr}}
\affiliation{\small Department of Medical Imaging Techniques,
Zonguldak Bülent Ecevit University, 67100, Zonguldak, Türkiye.\\}

\date{\today}

\begin{abstract}

We explore the sensitivity of future muon colliders to CP-violating interactions in the Higgs sector, specifically focusing on the process $\mu^- \mu^+ \to h \bar{\nu_{l}} \nu_{l} \to b\bar{b} \bar{\nu_{l}}\nu_{l}$. Using a model-independent approach within the framework of the Standard Model Effective Field Theory (SMEFT), we analyze the contribution of dimension-six operators to Higgs-gauge boson couplings, emphasizing CP-violating effects. To simulate the process, all signal and background events are generated through MadGraph. The analysis provides 95\% confidence level limits on the relevant Wilson coefficients $\tilde{c}_{HB}$, $\tilde{c}_{HW}$, $\tilde{c}_{\gamma}$, with a comparative discussion of existing experimental and phenomenological constraints. Our best constraints on the $\tilde{c}_{HB}$, $\tilde{c}_{HW}$, $\tilde{c}_{\gamma}$ with an integrated luminosity of 10 ab$^{-1}$ are $[-0.017428;0.018991]$, $[-0.002880;0.002586]$ and $[-0.010784;0.011381]$, respectively. In this context, this study highlights the capability of future muon collider experiments to probe new physics in the Higgs sector, potentially offering tighter constraints on CP-violating Higgs-gauge boson interactions than those provided by current colliders.

\end{abstract}


\vspace{5mm}

\maketitle


\section{Introduction}

The discovery of the 125 GeV Higgs boson by the ATLAS \cite{Aad:2012les} and CMS \cite{Chatrchyan:2012les} collaborations at the Large Hadron Collider (LHC) marked a turning point in our understanding of fundamental physics, confirming the Higgs mechanism as responsible for electroweak symmetry breaking (EWSB). This discovery was a key validation of the Standard Model (SM), yet it also opened new questions about the deeper nature of the Higgs boson. While the current experimental data align well with the SM predictions, particularly indicating that the Higgs is a CP-even scalar \cite{Aad:2016saf,Aad:2020czq,Aad:2020jkn,Sirunyan:2020wxc,Tumasyan:2022ssx}, the matter-antimatter asymmetry observed in the universe suggests there may be new, unexplored sources of CP violation \cite{Steigman:1976ghb,Cohen:1987efv,Steigman:2008ujb}.

In the SM, CP violation occurs primarily through the Cabibbo-Kobayashi-Maskawa (CKM) matrix in weak interactions \cite{Cabibbo:1963les,Kobayashi:1973hqw}. However, the level of CP violation it provides is insufficient to explain the baryon asymmetry of the universe, indicating that additional mechanisms must exist beyond the SM \cite{Riotto:1999jdx}. This motivates the exploration of Higgs boson interactions with gauge bosons and fermions, which could harbor new CP-violating effects. These interactions, if discovered, could provide key insights into the origin of the universe's matter dominance and guide us towards a more complete theory of particle physics \cite{Kuzmin:1985yla}.

One of the most promising frameworks for investigating these interactions is the Standard Model Effective Field Theory (SMEFT). By incorporating higher-dimensional operators, SMEFT allows for the systematic study of potential new physics contributions to Higgs couplings. In particular, CP-violating operators that couple the Higgs boson to gauge boson pairs (such as $HZ\gamma$, $HZZ$ and $HWW$) are of special interest. Precision measurements of these couplings could reveal deviations from SM predictions, pointing towards new physics.

Future high-energy colliders like the future muon collider are poised to make significant contributions in this area. With its ability to probe Higgs boson properties with unparalleled precision at multiple energy stages, the muon collider offers a unique opportunity to explore both CP-conserving and CP-violating interactions. By conducting detailed studies on the Higgs boson’s couplings with SM particles, the muon collider could potentially provide the experimental evidence for new sources of CP violation, offering a crucial key to solving the puzzle of the matter-antimatter asymmetry and leading to discoveries beyond the SM.

This study, therefore, focuses on exploring the potential CP-violating Higgs-gauge boson couplings using dimension-six operators within the SMEFT framework, and outlines the significance of these investigations in future collider experiments. In the following, after introducing our theoretical framework for the CP-violating Higgs-gauge boson couplings (Sec. II), we will present our strategy for testing for CP-violating Higgs-gauge boson couplings at the future muon collider in Sec. III, and our results in Sec. IV. Finally, we offer our conclusions in Sec. V.

\section{Effective Theory Approach}

Effective field theory (EFT) is a model-independent approach that simply analyzes deviations from the SM to study the properties of the Higgs boson. In this approach, new physics contributions beyond the SM, in addition to the SM Lagrangian, are parameterized as higher-dimensional operators. The effective Lagrangian respects the $SU(3)_C \times SU(2)_L \times U(1)_Y$ gauge symmetries, and the operators can be constrained separately by the Wilson coefficients as free parameters.

We consider the interactions of the Higgs boson and electroweak gauge bosons in the Strongly Interacting Light Higgs (SILH) basis \cite{Giudice:2007ops}. If the effects of physics beyond the SM are described by the dimension-six operators ${\cal O}_i$, the effective Lagrangian is given by:

\begin{eqnarray}
\label{eq.1} 
{\cal L}={\cal L}_{\text{SM}}+\sum_{i}{\overline{c}_i}{\cal O}_i={\cal L}_{\text{SM}}+{\cal L}_{\text{CPV}}
\end{eqnarray}

{\raggedright where ${\overline{c}_i}$ are dimensionless Wilson coefficients normalized by the form of the new physics scale $\Lambda$ identified with the $W$-boson mass $m_W$ and ${\cal L}_{\text{SM}}$ is the dimension-four SM Lagrangian. In this paper, we focus on the CP-violating interactions of the Higgs and electroweak gauge bosons, and these are written in terms of the SILH basis:}

\begin{eqnarray}
\label{eq.2} 
{\cal L}_{\text{CPV}}=\frac{ig\tilde{c}_{HW}}{m_W^2}D^\mu \Phi^\dagger T_{2k} {D}^\nu\Phi \widetilde{W}_{\mu\nu}^k+\frac{ig^\prime \tilde{c}_{HB}}{m_W^2}D^\mu \Phi^\dagger {D}^\nu\Phi \widetilde{B}_{\mu\nu}+\frac{g^{\prime 2} \tilde{c}_{\gamma}}{m_W^2}\Phi^\dagger \Phi B_{\mu\nu} \widetilde{B}^{\mu\nu}
\end{eqnarray}

{\raggedright where $T_{2k}=\sigma_k/2$ with $\sigma_k$ is the Pauli matrices. ${B}_{\mu\nu}=\partial_\mu B_\nu - \partial_\nu B_\mu$ and ${W}_{\mu\nu}^k=\partial_\mu W_\nu^k - \partial_\nu W_\mu^k + g\epsilon_{ij}^k W_\mu^i W_\nu^j$ are the field strength tensors corresponding to $U(1)_Y$ and $SU(2)_L$ of the SM gauge groups, respectively, with gauge coupling constants $g^\prime$ and $g$. $\widetilde{B}_{\mu\nu}=\frac{1}{2}\epsilon_{\mu\nu\rho\sigma}B^{\rho\sigma}$ and $\widetilde{W}_{\mu\nu}^k=\frac{1}{2}\epsilon_{\mu\nu\rho\sigma}W^{\rho\sigma k}$ are the dual field strength tensors. $D^\mu$ is covariant derivative operator and $\Phi$ is the Higgs doublet in the SM.}

An effective Lagrangian in the mass basis with anomalous Higgs couplings is used for a phenomenological and experimental approach. It has proven to be a useful approach to relate experimental bounds expressed in terms of anomalous couplings to phenomenological bounds obtained by theories or models. The relevant subset of anomalous $HZ\gamma$, $HZZ$ and $HWW$ couplings in the mass basis and in the unitary gauge is written as \cite{Alloul:2014hws}: 

\begin{eqnarray}
\label{eq.3} 
{\cal L}=-\frac{1}{4}\widetilde{g}_{hzz}Z_{\mu\nu}\widetilde{Z}^{\mu\nu}h-\frac{1}{2}\widetilde{g}_{h\gamma z}Z_{\mu\nu}\widetilde{F}^{\mu\nu}h-\frac{1}{2}\widetilde{g}_{hww}W^{\mu\nu}\widetilde{W}^\dagger_{\mu\nu}h
\end{eqnarray} 

{\raggedright where $h$ is Higgs boson field. The relation between these anomalous coupling coefficients in the mass basis and the dimension-six coefficients is given by:}

\begin{eqnarray}
\label{eq.4} 
\widetilde{g}_{hzz}=\frac{2g}{c_W^2 m_W}\left[\tilde{c}_{HB}s_W^2-4\tilde{c}_{\gamma}s_W^4+c_W^2\tilde{c}_{HW}\right]
\end{eqnarray}

\begin{eqnarray}
\label{eq.5} 
\widetilde{g}_{h\gamma z}=\frac{gs_W}{c_W m_W}\left[\tilde{c}_{HW}-\tilde{c}_{HB}+8\tilde{c}_{\gamma}s_W^2\right]
\end{eqnarray}

\begin{eqnarray}
\label{eq.6} 
\widetilde{g}_{hww}=\frac{2g}{m_W}\tilde{c}_{HW}
\end{eqnarray}

{\raggedright where $s_W=\text{sin}\theta_W$ and $c_W=\text{cos}\theta_W$ with $\theta_W$ being the weak mixing angle. There are three Wilson coefficients, $\tilde{c}_{HB}$, $\tilde{c}_{HW}$ and $\tilde{c}_\gamma$ for CP-violating couplings.}

We focus on the sensitivity study of $\tilde{c}_{HB}$, $\tilde{c}_{HW}$ and $\tilde{c}_\gamma$ coefficients in the anomalous $HZ\gamma$, $HZZ$ and $HWW$ vertices through the $\mu^- \mu^+ \to h \bar{\nu_{l}} \nu_{l} \to b\bar{b} \bar{\nu_{l}}\nu_{l}$ process with single Higgs boson production at the muon collider, and the investigation of the CP-violating properties of the anomalous couplings. In this paper, analysis of the dimension-six operators in Higgs-gauge boson couplings are performed into {\sc MadGraph5}$\_$aMC@NLO \cite{Alwall:2014cvc} based on Monte Carlo simulations using FeynRules  \cite{Alloul:2014tfc} and the UFO \cite{Degrande:2012acs} framework. The HEL model file \cite{Alloul:2014hws} containing the 39 dimension-six operators and their corresponding Wilson coefficients is implemented in FeynRules.

There are many phenomenological studies to obtain constraints on the Wilson coefficients of CP-conserving and/or CP-violating dimension-six operators using various channels at the $pp$ \cite{Ellis:2015tdw,Englert:2016edw,Khanpour:2017ssa,Ferreira:2017qmj,Denizli:2019oxc,Denizli:2021uhb,Denizli:2021pkw}, $ee$ \cite{Ellis:2014opx,Kumar:2015rvx,Ellis:2016dmk,Khanpour:2017ubw,Alam:2017rma,Ellis:2017klz,Denizli:2018rca,Karadeniz:2020yvz}, $ep$ \cite{Kuday:2018yza,Hesari:2018ygv} and $\mu\mu$ \cite{Spor:2024sdk} colliders. Focusing on the design of a muon collider with high luminosity and high energy, the International Muon Collider Collaboration (IMCC) \cite{Accettura:2023tls} has recently been studying the possibility of developing a muon collider with 10 TeV center-of-mass energy. In this study, we consider a 10 TeV center-of-mass energy ($\sqrt{s}=10$ TeV) muon collider that produces an integrated luminosity equal to 10 ab$^{-1}$ per interaction point (IP). Lepton colliders are generally known to have much smaller backgrounds and a cleaner environment than hadron colliders. Because muons are 207 times heavier than electrons, muon collisions produce less synchrotron radiation than electron-positron collisions, making it easier to accelerate them to high energies with a circular collider. Muon colliders, which have the advantage of both being lepton collider and operating at multi-TeV energies, offer a unique opportunity in the search for new physics by enabling precise measurements of Higgs-gauge bosons couplings.

\begin{figure}[H]
\centerline{\scalebox{0.35}{\includegraphics{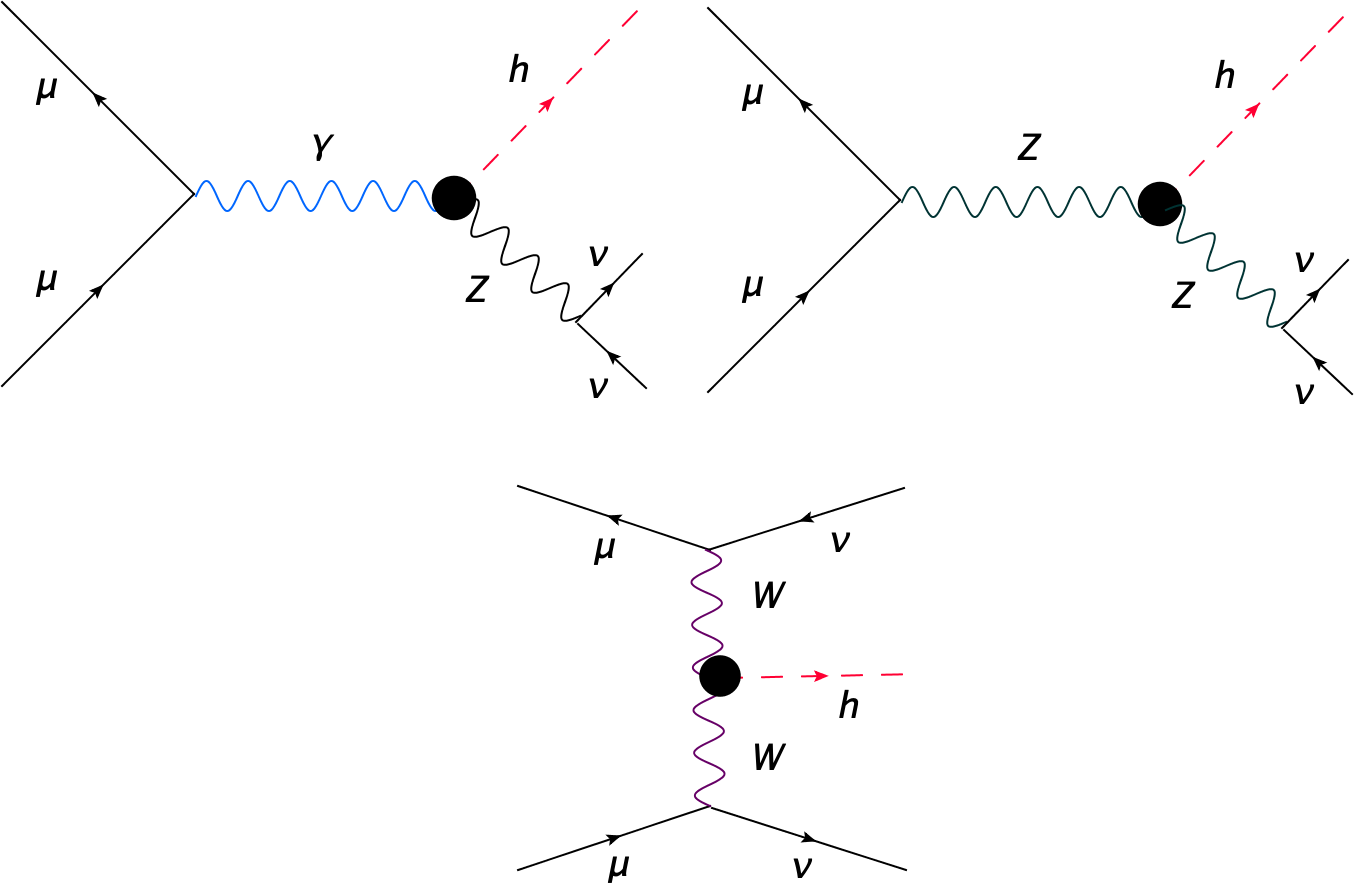}}}
\caption{ \label{fig:1}  Feynman diagrams of the $\mu^- \mu^+ \to h \bar{\nu_{l}} \nu_{l}$ process including the anomalous Higgs-gauge bosons vertices.}
\end{figure}

The Feynman diagrams of the $\mu^- \mu^+ \to h \bar{\nu_{l}} \nu_{l}$ process including the anomalous Higgs-gauge bosons vertices are given in Fig.~\ref{fig:1}. These Feynman diagrams represent signal processes involving new physics contributions from the anomalous $HZ\gamma$, $HZZ$ and $HWW$ vertices.

\begin{figure}[H]
\centerline{\scalebox{1.31}{\includegraphics{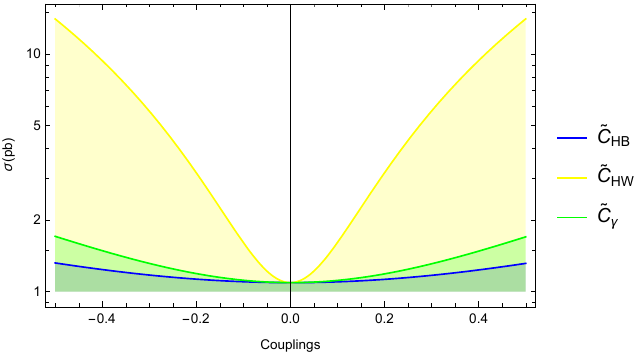}}}
\caption{ \label{fig:2} The total cross-sections of the process $\mu^- \mu^+ \to h \bar{\nu_{l}} \nu_{l} \to b\bar{b} \bar{\nu_{l}}\nu_{l}$ in terms of the coefficients $\tilde{c}_{HB}$, $\tilde{c}_{HW}$ and $\tilde{c}_{\gamma}$ at the muon collider.}
\end{figure}

The total cross-sections of the process $\mu^- \mu^+ \to h \bar{\nu_{l}} \nu_{l} \to b\bar{b} \bar{\nu_{l}}\nu_{l}$  in term of coefficients $\tilde{c}_{HB}$, $\tilde{c}_{HW}$ and $\tilde{c}_{\gamma}$ at the muon collider are shown in Fig.~\ref{fig:2}. The calculation method of the total cross-sections in Fig.~\ref{fig:2} is that a certain coefficient is variable each time, while the other coefficients are fixed to zero. All Wilson coefficients being equal to zero ($\tilde{c}_{HB}=\tilde{c}_{HW}=\tilde{c}_{\gamma}=0$) correspond to the cross-section of the SM contribution in the process $\mu^- \mu^+ \to h \bar{\nu_{l}} \nu_{l} \to b\bar{b} \bar{\nu_{l}}\nu_{l}$.

\section{Event Generation and Cut-Based Analysis}

In this section, we describe our simulation setup and apply a cut-based analysis to derive constraints on anomalous $HZ\gamma$, $HZZ$ and $HWW$ Higgs-gauge boson couplings via the process $\mu^- \mu^+ \to h \bar{\nu_{l}} \nu_{l} \to b\bar{b} \bar{\nu_{l}}\nu_{l}$ at the muon collider. The process $\mu^- \mu^+ \to h \bar{\nu_{l}} \nu_{l} \to b\bar{b} \bar{\nu_{l}}\nu_{l}$ serves as signal, including SM contributions, pure new physics terms with contributions from non-zero coefficients $\tilde{c}_{HB}$, $\tilde{c}_{HW}$, $\tilde{c}_{\gamma}$ and the interference terms between the SM and pure new physics terms. On the other hand, some background processes are examined to have a more realistic simulation. The background process is defined as follows: the $B_{H\nu\nu}$ represents the SM background process $\mu^- \mu^+ \to h \bar{\nu_{l}} \nu_{l} \to b \bar{b} \bar{\nu_{l}} \nu_{l}$, which has the same final state as the signal process. The following other backgrounds are: i) the process $\mu^- \mu^+ \to b\bar{b} \bar{\nu_{l}}\nu_{l}$ is labeled as $B_{b\bar{b}\nu\bar{\nu}}$, considering the full contribution of the $b\bar{b} \bar{\nu_{l}}\nu_{l}$ final state. ii) the pair production of top quark is realized through the process $\mu^- \mu^+ \to t\bar{t} \to W^+ b W^- \bar{b} \to \ell^+ {\nu_{l}} b \ell^- \bar{\nu}_{l} \bar{b}$ labeling as $B_{t\bar{t}}$, where one top quark (anti-top quark) decays to $W^+b$ ($W^-\bar{b}$) which contains the leptonic decay channel of the $W^\pm$-boson.

\begin{figure}[H]
\centerline{\scalebox{0.50}{\includegraphics{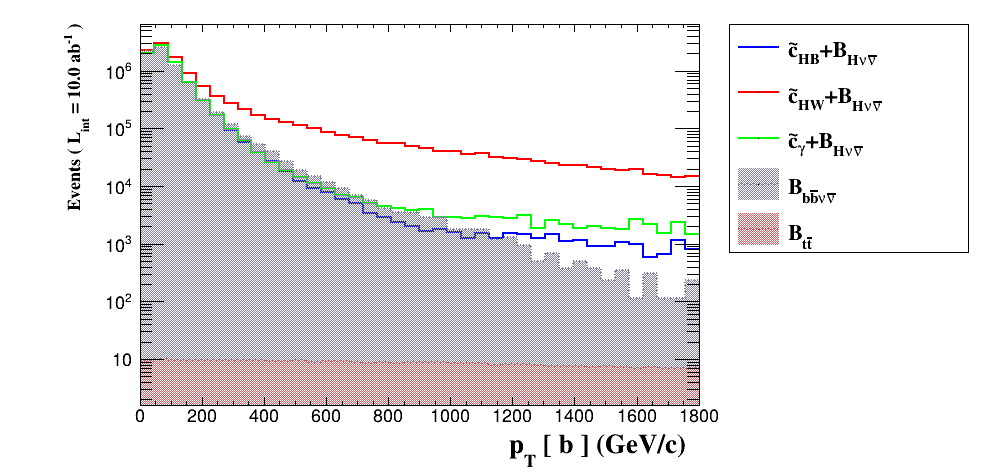}}}
\caption{ \label{fig:3}  The number of events as a function of $p^{b}_{T}$ for the process $\mu^- \mu^+ \to h \bar{\nu_{l}} \nu_{l} \to b\bar{b} \bar{\nu_{l}}\nu_{l}$ and related backgrounds at muon collider with $\sqrt{s} = 10$ TeV.}
\end{figure}

\begin{figure}[H]
\centerline{\scalebox{0.50}{\includegraphics{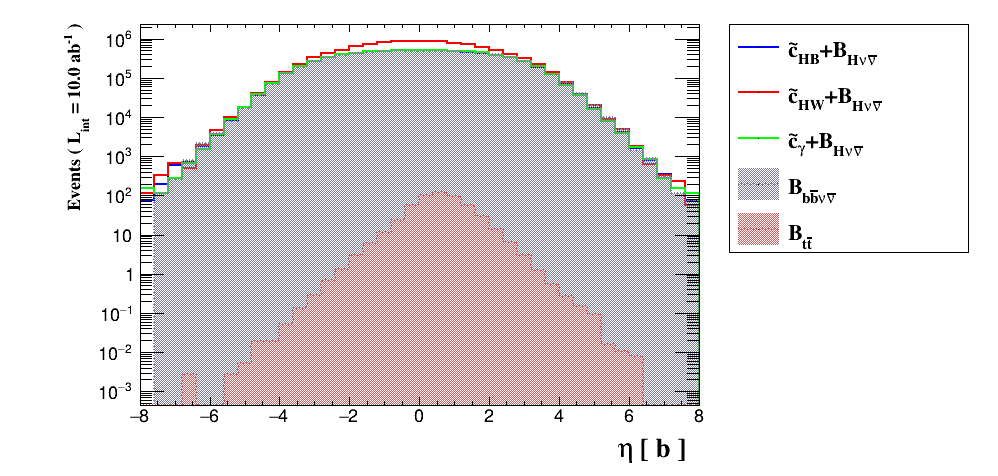}}}
\caption{ \label{fig:4}  The number of events as a function of $\eta^{b}$ for the process $\mu^- \mu^+ \to h \bar{\nu_{l}} \nu_{l} \to b\bar{b} \bar{\nu_{l}}\nu_{l}$ and related backgrounds at muon collider with $\sqrt{s} = 10$ TeV.}
\end{figure}

To simulate signals for each coupling and relevant backgrounds, we generated 400 k samples using {\sc MadGraph5}$\_$aMC@NLO. For signal identification, a set of cuts was applied to separate signal and backgrounds in the $b\bar{b}\nu\nu$ final state. First the transverse momentum distributions for the final state b-quark for signal and relevant background  processes at muon collider are presented in Fig.~\ref{fig:3}. It can be seen from Fig.~\ref{fig:3} that the signal can be separated from the backgrounds by $p^{b}_{T} > 50$ GeV. Fig.~\ref{fig:4} shows pseudo-rapidity distributions for b-quark, with a distinct separation in signal occurring for $|\eta^{b}| < 2.4$, which is implemented as Cut-1.

\begin{figure}[H]
\centerline{\scalebox{0.50}{\includegraphics{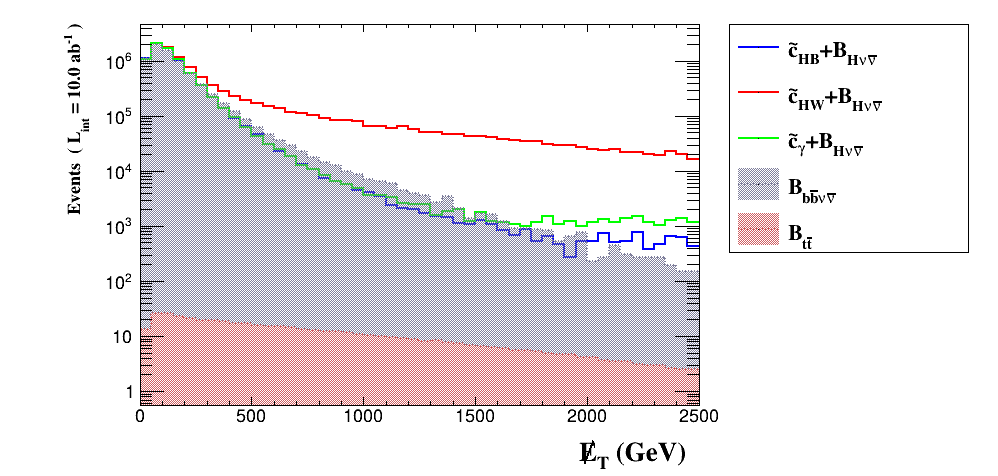}}}
\caption{ \label{fig:5}  The number of events as a function of $\slashed{E}_{T}$ for the process $\mu^- \mu^+ \to h \bar{\nu_{l}} \nu_{l} \to b\bar{b} \bar{\nu_{l}}\nu_{l}$ and related backgrounds at muon collider with $\sqrt{s} = 10$ TeV.}
\end{figure}

\begin{figure}[H]
\centerline{\scalebox{0.50}{\includegraphics{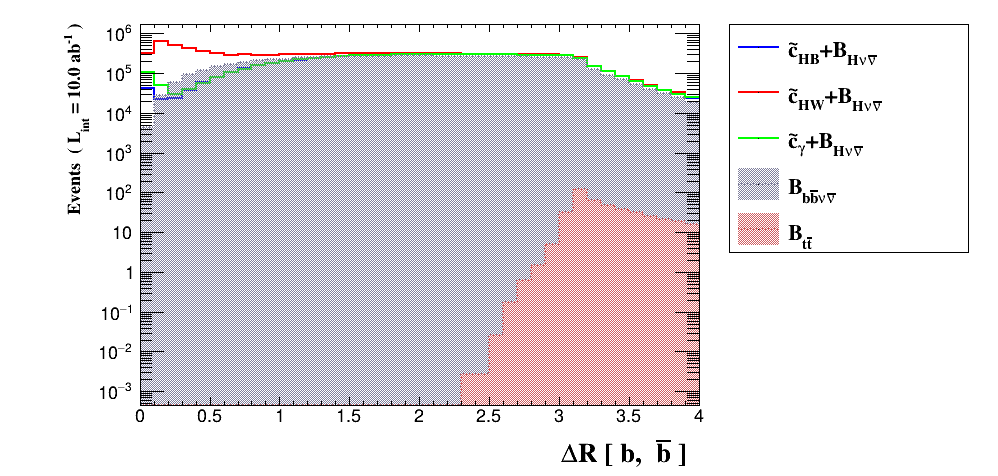}}}
\caption{ \label{fig:6}  The number of events as a function of $\Delta{R}_{b\bar{b}}$ for the process $\mu^- \mu^+ \to h \bar{\nu_{l}} \nu_{l} \to b\bar{b} \bar{\nu_{l}}\nu_{l}$ and related backgrounds at muon collider with $\sqrt{s} = 10$ TeV.}
\end{figure}

On the other hand, Fig.~\ref{fig:5} and Fig.~\ref{fig:6} demonstrate that background suppression is further achieved by imposing missing transverse energy $\slashed{E}_T > 100$ (Cut-2) and restricting the distance between final state b-quarks $\Delta R(b,\bar{b}) < 1.6$ (Cut-3).

\begin{figure}[H]
\centerline{\scalebox{0.55}{\includegraphics{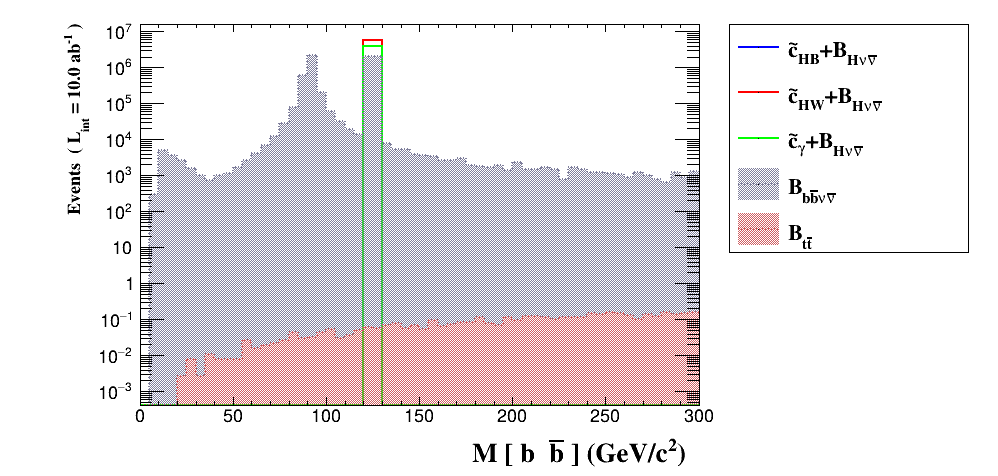}}}
\caption{ \label{fig:7} The number of events as a function of $M_{b\bar{b}}$ for the process $\mu^- \mu^+ \to h \bar{\nu_{l}} \nu_{l} \to b\bar{b} \bar{\nu_{l}}\nu_{l}$ and related backgrounds at muon collider with $\sqrt{s} = 10$ TeV.}
\end{figure}

For the invariant mass of the final state b-quarks from the decay of the Higgs boson, shown in Fig.~\ref{fig:7}, we set $120<M_{b\bar{b}}<130$ GeV (Cut-4). 

Finally, the list of particle-level cuts that are used in our calculations is given in a flow in Table~\ref{tab1}. On the other hand, Table~\ref{tab2} shows the number of events after each cut for the signals ($\tilde{c}_{HB}=0.1$, $\tilde{c}_{HW}=0.1$, $\tilde{c}_{\gamma}=0.1$) and relevant backgrounds. The number of events are calculated with an integrated luminosity of \( \mathcal{L}_{\text{int}} = 10 \) ab\(^{-1}\) for each cut of the signal and relevant backgrounds. As seen in Table~\ref{tab2}, Cut-4 shows a significant improvement for enhancing separation between the signals and related backgrounds.

\begin{table}[h!]
\centering
\caption{Particle-level selection cuts for the $\mu^- \mu^+ \to h \bar{\nu_{l}} \nu_{l} \to b\bar{b} \bar{\nu_{l}}\nu_{l}$ signal at the muon collider.}
\label{tab1}
\begin{tabular}{p{0.4\textwidth} p{0.6\textwidth}}
\hline
\hline
Kinematic Cuts & $\tilde{c}_{HB}$, $\tilde{c}_{HW}$, $\tilde{c}_{\gamma}$  \\
\hline
Cut-1   & $p^b_T > 50$ , $|\eta^{b}| < 2.4$   \\
\hline
Cut-2   & $\slashed{E}_T > 100$ GeV \\
\hline
Cut-3   & $\Delta R(b,\bar{b}) < 1.6$\\
\hline
Cut-4   & $120<M_{b\bar{b}}<130$ GeV\\
\hline
\end{tabular}
\end{table}

\begin{table}[H]
\centering
\caption{The number of events of Higgs-gauge boson couplings for the process $\mu^- \mu^+ \to h \bar{\nu_{l}} \nu_{l} \to b\bar{b} \bar{\nu_{l}}\nu_{l}$ and relevant backgrounds. Here, $\epsilon[\%]$ is the relative efficiency of each cut and M = $10^{6}$ and the $b$-tagging efficiency $\varepsilon_b = 0.7$ is considered.}
\label{tab2}
\begin{tabular}{p{2.2cm}p{1.5cm}p{1cm}p{1.5cm}p{1cm}p{1.5cm}p{1cm}p{1.5cm}p{1cm}p{1.5cm}p{1cm}}
\hline \hline
MuCol. & \multicolumn{2}{c}{Presel.} & \multicolumn{2}{c}{Cut-1} & \multicolumn{2}{c}{Cut-2} & \multicolumn{2}{c}{Cut-3} & \multicolumn{2}{c}{Cut-4} \\ \hline
Signal & Events & $\epsilon$[\%] & Events & $\epsilon$[\%] & Events & $\epsilon$[\%] & Events & $\epsilon$[\%] & Events & $\epsilon$[\%]\\ \hline
$\tilde{c}_{HB}=0.1$ & 7.72M & --- & 3.99M & 52 & 2.73M & 68 & 1.76M & 64 & 1.76M & 100\\
$\tilde{c}_{HW}=0.1$ & 11.28M & --- & 6.96M & 61 & 5.71M & 82 & 4.66M & 81 & 4.66M & 100\\
$\tilde{c}_{\gamma}=0.1$ & 7.83M & --- & 4.09M & 52 & 2.85M & 69 & 1.87M & 66 & 1.87M & 100\\
\cline{1-1}
Backgrounds & & & & & \\ \cline{1-1}
$B_{b \bar{b}\bar{\nu}_l\nu_l}$ & 7.66M & --- & 3.77M & 49 & 2.69M & 71 & 1.95M & 72 & 0.95M & 49\\
$B_{t\bar{t}}$ & 534 & --- & 513 & 96 & 478 & 93 & 0 & 0 & 0 & ---\\
\hline \hline
\end{tabular}
\end{table}

\begin{figure}[H]
\centerline{\scalebox{0.38}{\includegraphics{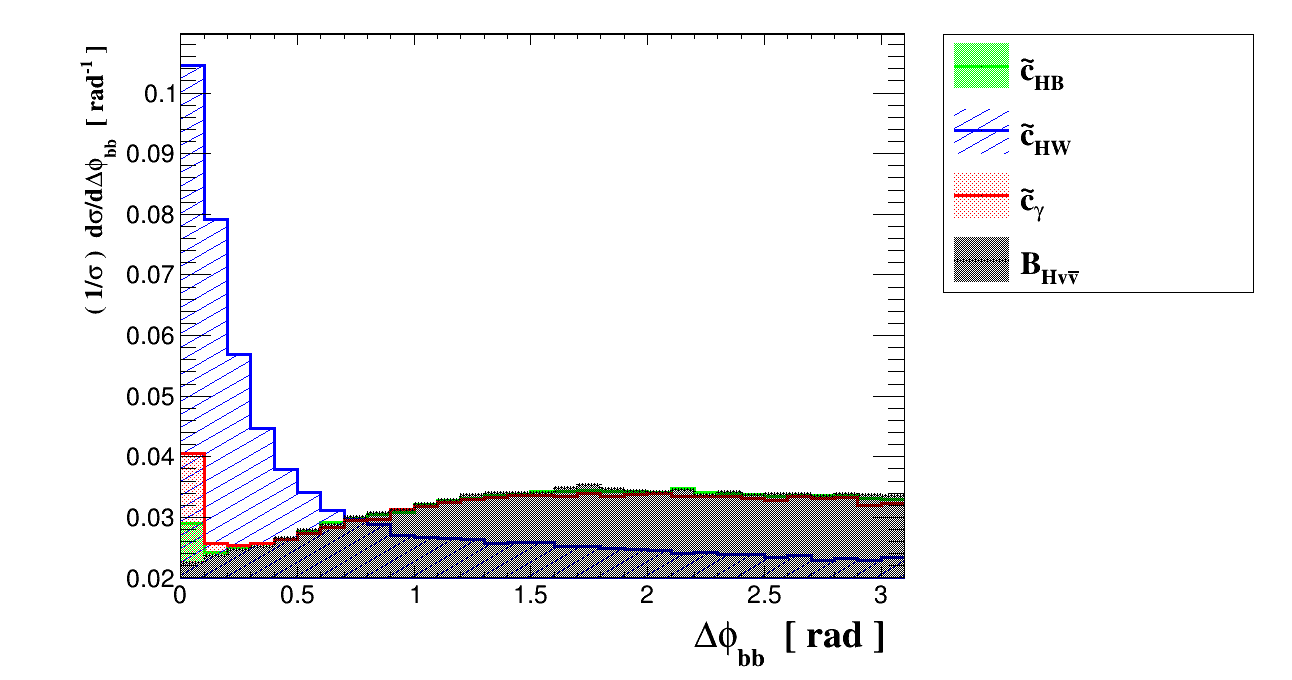}}}
\caption{ \label{fig:8}  The number of events as a function of $\Delta{\phi}_{b\bar{b}}$ for the process $\mu^- \mu^+ \to h \bar{\nu_{l}} \nu_{l} \to b\bar{b} \bar{\nu_{l}}\nu_{l}$ and SM background at muon collider with $\sqrt{s} = 10$ TeV.}
\end{figure}

In order to investigate the CP-violating nature of the Higgs-gauge boson interactions, we analyze the differential distribution of the azimuthal angle difference between the two $b$-jets, $\Delta \phi_{b\bar{b}}$. As shown in Fig.~\ref{fig:8}, the distribution exhibits characteristic modulations depending on the presence of CP-violating operators. Specifically, scenarios involving the CP-odd couplings $\tilde{c}_{HB}$, $\tilde{c}_{HW}$, and $\tilde{c}_{\gamma}$ show noticeable deviations from the SM background ($B_{H\nu\bar{\nu}}$), particularly at lower values of $\Delta \phi_{b\bar{b}}$. These deviations are indicative of interference effects arising from CP-odd interactions and demonstrate the potential of angular observables in distinguishing CP-violating new physics from both the SM and CP-conserving dimension-six operators.

\section{Expected Sensitivities on Dim-6 CP-Violating Higgs-gauge Boson Couplings}

The sensitivities of dimension-six Higgs-gauge boson interactions to anomalous couplings was examined using simulated data from muon collisions at a center-of-mass energy of 10 TeV. The analysis was conducted using a $\chi^2$ test to quantify the deviations from SM predictions. This test compares the total cross-section, including contributions from effective couplings, with the cross-section of backgrounds. Both systematic and statistical uncertainties were considered to obtain accurate error margins. The $\chi^{2}$ test is defined as follows:

\begin{equation}
\chi^2(\tilde{c}_{HB}, \tilde{c}_{HW}, \tilde{c}_{\gamma})=\Biggl(\frac{\sigma_{SM}(\sqrt{s})-\sigma_{Total}(\sqrt{s}, \tilde{c}_{HB}, \tilde{c}_{HW}, \tilde{c}_{\gamma}}
{\sigma_{B}(\sqrt{s})\sqrt{(\delta_{sys})^2 + (\delta_{st})^2}}\Biggr)^2,
\end{equation}

Here, $\sigma_{SM}(\sqrt{s})$ represents the cross-section of the SM background, and $\sigma_{total}(\sqrt{s}, \tilde{c}_{HB}, \tilde{c}_{HW}, \tilde{c}_{\gamma})$ is the total cross-section of both the new physics coming from beyond the SM and the SM. On the other hand, $\sigma_{B}(\sqrt{s})$ is the total background that we consider in the analysis. $\delta_{st}=\frac{1}{\sqrt{N_{B}}}$ and $\delta_{sys}$ are the statistical error and systematic uncertainty, respectively. The event number of the cumulative background is defined as $N_{B}={\cal L}\times \sigma_{B}\times\varepsilon_b$, where ${\cal L}$ is the integrated luminosity and $\varepsilon_b$ is the $b$-tagging efficiency.

Identifying jets originating from $b$-quarks ($b$-tagging) is crucial in collider-based particle physics experiments. It plays a significant role in studies involving Higgs boson decays to $b\bar{b}$, Higgs boson production along with the top-quark pair, precision measurements in the SM, and searches for physics beyond the SM \cite{Aad:2019ghb}. In experimental analyses, $b$-tagging algorithms exploit properties such as the relatively long lifetime, large mass, and high decay multiplicity of $b$-hadrons. The performance of $b$-tagging is typically characterized by two parameters: the $b$-tagging efficiency ($\varepsilon_b$), which is the probability of correctly identifying a $b$-jet, and the misidentification rates of charm ($c$) or light-flavor ($u$, $d$, $s$ quarks, or gluons) jets as $b$-jets. These rates depend on the specific algorithm employed and the kinematic properties, such as the transverse momentum of the jet. This study adopts a constant $b$-tagging efficiency of $\varepsilon_b = 0.7$, consistent with values used in recent phenomenological analyses~\cite{Sun:2023tlm,Liu:2021pas}.

The primary focus of the analysis was to obtain the sensitivities on various Higgs-gauge boson couplings with the at 95\% Confidence Level (C.L.). Systematic errors were accounted for in the calculations to reflect more realistic experimental conditions. Here, possible sources of systematic uncertainties are integrated luminosities, photon efficiencies, jet-photon misidentification, detector efficiency and background estimation. The obtained sensitivities on Higgs-gauge couplings at 95\% C.L. are given in  Table~\ref{tab3} under the systematic uncertainties of 0\%, 3\% and 5\% and compared with the experimental results.

\begin{table}[H]
\centering
\caption{Sensitivities at $95\%$ C.L. on the anomalous Higgs-gauge boson couplings for various systematic uncertainties.}
\label{tab3}
\begin{tabular}{ccc c}
\hline
\hline
\multicolumn{3}{c}{} & \multicolumn{1}{c}{Muon Collider} \\
\hline
Effective Couplings & Experimental Results (ATLAS) & Systematic Errors & Our Projection \\
\hline
& $[-0.23;0.23]$ \cite{Aad:2016hws} & $\delta=0\%$ & $[-0.017428;0.018991]$  \\
$\tilde{c}_{HB}$  & $[-0.16;0.16]$ \cite{Aaboud:2018yer} & $\delta=3\%$ &$[-0.116988;0.118550]$  \\
& $[-0.065;0.063]$ \cite{ATLAS:2019sdf} & $\delta=5\%$ & $[-0.151243;0.152805]$ \\
\hline
& $[-0.23;0.23]$ \cite{Aad:2016hws} & $\delta=0\%$ &$[-0.002880;0.002586]$  \\
$\tilde{c}_{HW}$ & $[-0.16;0.16]$ \cite{Aaboud:2018yer} & $\delta=3\%$ & $[-0.017882;0.017588]$  \\
& $[-0.065;0.063]$ \cite{ATLAS:2019sdf} & $\delta=5\%$ & $[-0.023040;0.022746]$  \\
\hline
& $[-0.0018;0.0018]$ \cite{Aad:2016hws} & $\delta=0\%$ & $[-0.010784;0.011381]$  \\
$\tilde{c}_{\gamma}$ &  $[-0.00028;0.00043]$ \cite{ATLAS:2019sdf} & $\delta=3\%$ &$[-0.071416;0.072012]$  \\ 
& &  $\delta=5\%$ & $[-0.092276;0.092872]$ \\
\hline
\end{tabular}
\end{table}

The results in Table~\ref{tab3} indicate that our projections provide particularly strong sensitivity to deviations from the SM, with the limits on the $\tilde{c}_{HB}$, $\tilde{c}_{HW}$ couplings being highly constrained at 10 TeV muon collider with integrated luminosities of 10 ab$^{-1}$ comparing with the ATLAS results.

\begin{figure}[H]
\centerline{\scalebox{1.7}{\includegraphics{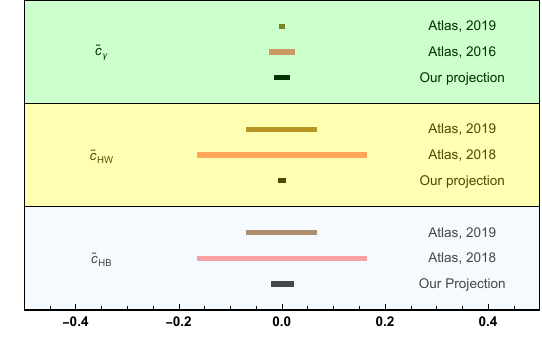}}}
\caption{ \label{fig:9} Comparison of ATLAS bounds and obtained sensitivities on the anomalous $\tilde{c}_{HB}$, $\tilde{c}_{HW}$, $\tilde{c}_{\gamma}$ parameters via the process $\mu^- \mu^+ \to h \bar{\nu_{l}} \nu_{l} \to b\bar{b} \bar{\nu_{l}}\nu_{l}$ at muon collider.}
\end{figure}

\begin{figure}[H]
\centering
\begin{subfigure}{0.5\linewidth}
\includegraphics[width=\linewidth]{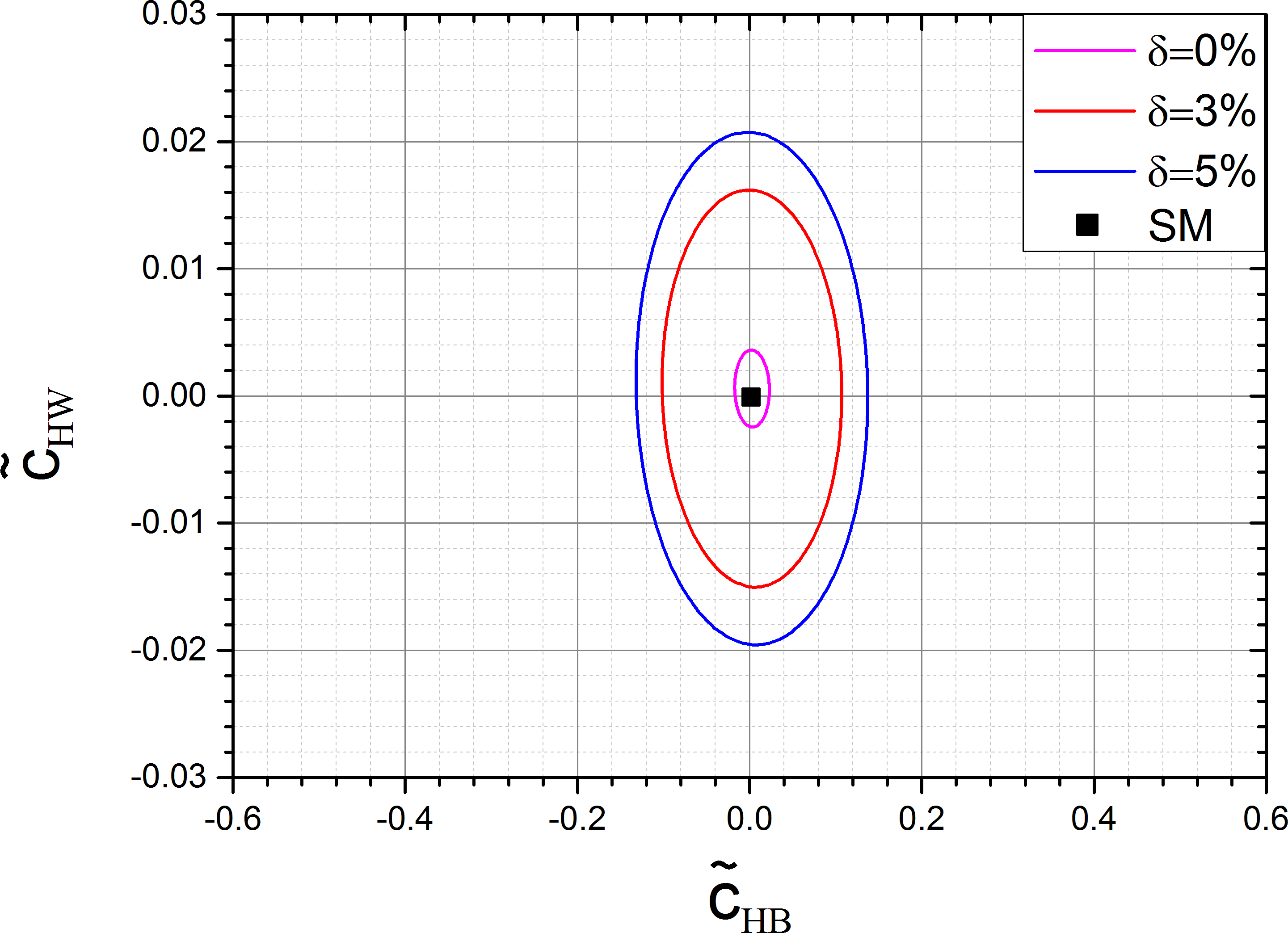}
\caption{}
\label{fig10:a}
\end{subfigure}\hfill
\begin{subfigure}{0.49\linewidth}
\includegraphics[width=\linewidth]{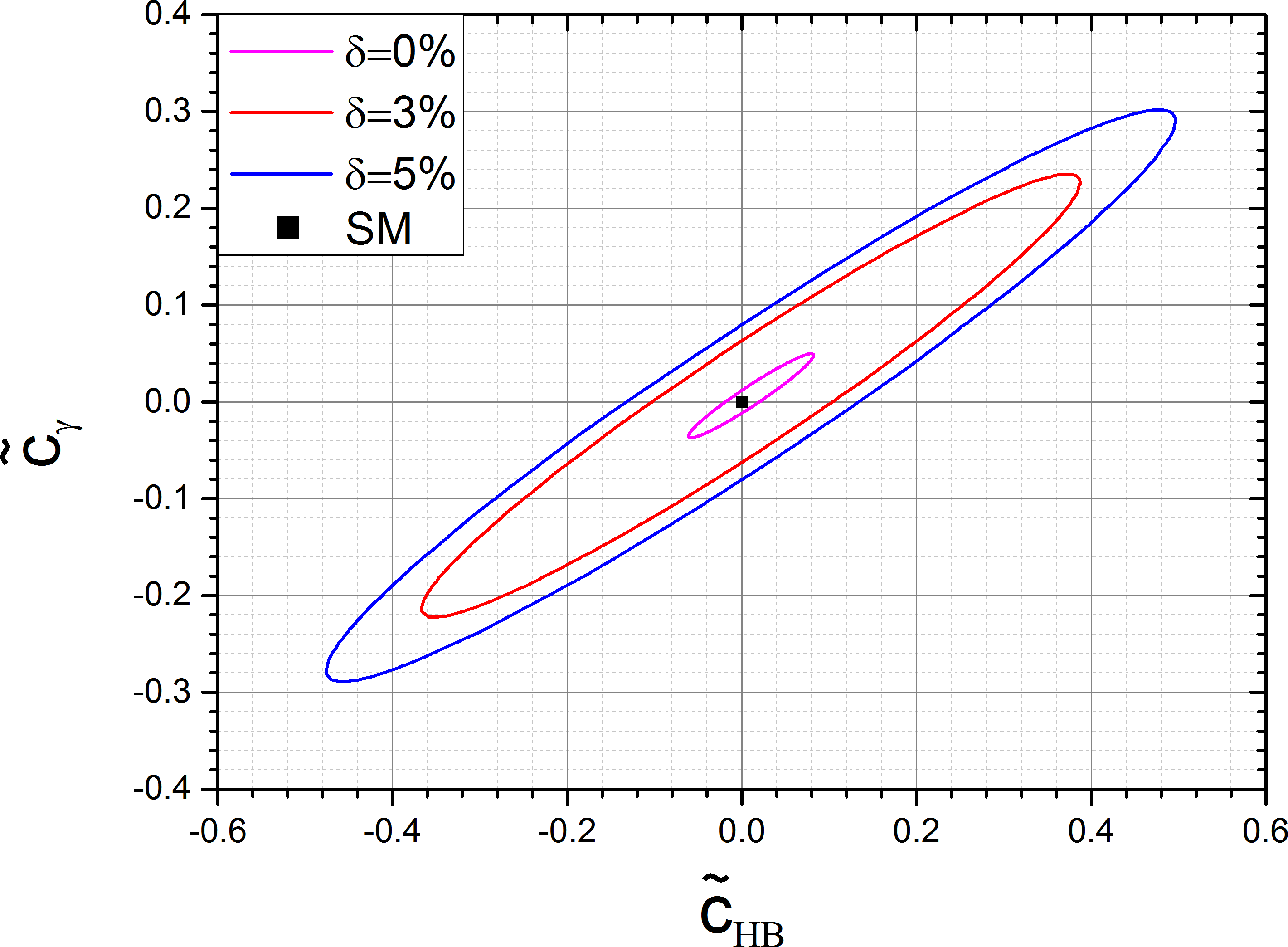}
\caption{}
\label{fig10:b}
\end{subfigure}\hfill
\begin{subfigure}{0.5\linewidth}
\includegraphics[width=\linewidth]{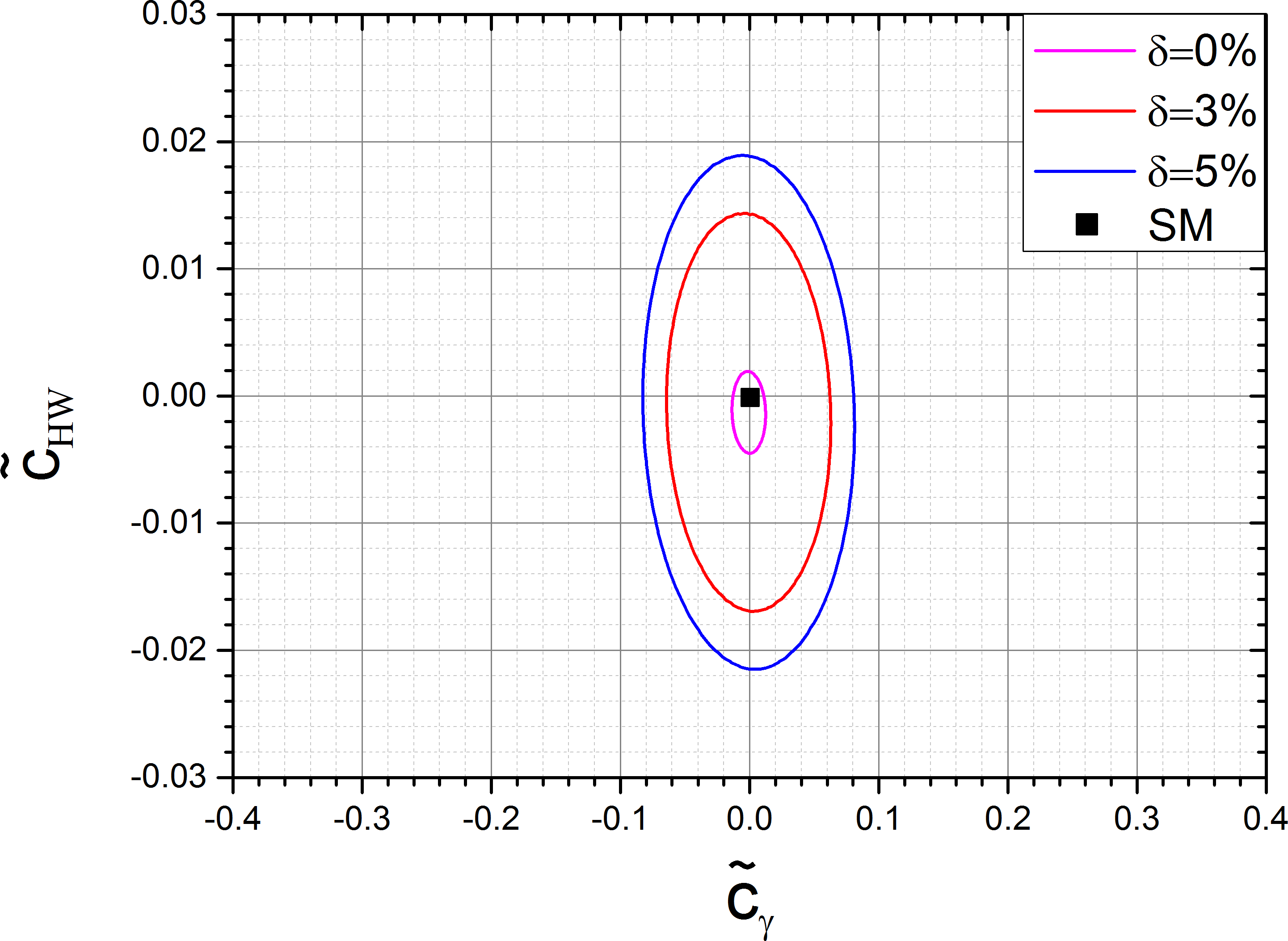}
\caption{}
\label{fig10:c}
\end{subfigure}\hfill

\caption{Two-dimensional 95\% C.L. intervals in plane for $\tilde{c}_{HB}-\tilde{c}_{HW}$ (a), $\tilde{c}_{HB}-\tilde{c}_{\gamma}$ (b) and $\tilde{c}_{\gamma}-\tilde{c}_{HW}$ (c) with taking systematic error $\delta_{sys}=0\%$, $3\%$ and $5\%$ at the muon collider. The black square dot symbol represents the SM expectation.}
\label{fig:10}
\end{figure}

On the other hand, Fig.~\ref{fig:9} compare the obtained sensitivities at future muon collider and the experimental results for ATLAS results with barcharts \cite{Aaboud:2018yer,ATLAS:2019sdf}. As seen in Fig.~\ref{fig:9} our results demonstrates the potential of future muon colliders to probe new physics scenarios with unprecedented precision. Finally, to evaluate how constraints on a specific CP-violating coupling are affected by the presence of others, two-dimensional contours are considered in which the other coefficient is fixed at zero while scanning is performed on these two Wilson coefficients. Fig.~\ref{fig:10} illustrates the 95\% C.L. contours in the $\tilde{c}_{HB}-\tilde{c}_{HW}$, $\tilde{c}_{HB}-\tilde{c}_{\gamma}$, and $\tilde{c}_{\gamma}-\tilde{c}_{HW}$ planes, based on a two-parameter analysis at muon collider for the systematic uncertainties of 0\%, 3\% and 5\%.

\section{Conclusions}

We have investigated the sensitivity of CP-violating dimension-six operators in the Higgs-gauge boson sector using the process $\mu^- \mu^+ \to h \bar{\nu_{l}} \nu_{l} \to b\bar{b} \bar{\nu_{l}}\nu_{l}$ at a future muon collider with $\sqrt{s} = 10$ TeV and an integrated luminosity of 10 ab$^{-1}$. Our analysis, which included a detailed study of kinematic distributions such as the transverse momentum, pseudo-rapidity and the $\Delta R$ of the final state $b$-quarks, missing energy transverse and the invariant mass of the Higgs boson, aimed to provide new limits on the Wilson coefficients $\tilde{c}_{HB}$, $\tilde{c}_{HW}$, and $\tilde{c}_{\gamma}$. The obtained sensitivities at 95\% C.L. for these parameters are $[-0.017428; 0.018991]$, $[-0.002880; 0.002586]$, and $[-0.010784; 0.011381]$, respectively. When compared to the experimental limits of $[-0.23;0.23]$\cite{Aad:2016hws}, $[-0.16;0.16]$ \cite{Aaboud:2018yer} and $[-0.065; 0.063]$ \cite{ATLAS:2019sdf} for both $\tilde{c}_{HB}$ and $\tilde{c}_{HW}$, and $[-0.00028; 0.00043]$ \cite{ATLAS:2019sdf} for $\tilde{c}_{\gamma}$, we observe a significant improvement in the constraints on the $\tilde{c}_{HB}$ and $\tilde{c}_{HW}$ couplings. However, for $\tilde{c}{\gamma}$, while our results do not surpass the experimental limits provided by ATLAS at $\sqrt{s} = 13$ TeV with an integrated luminosity of 139 fb$^{-1}$, they still offer valuable insights that are consistent with phenomenological study at CLIC and muon collider in Ref.~\cite{Spor:2024sdk}. Phenomenological study on CP-violating dimension-six operators through the process $e^+e^- \to H\nu\bar{\nu}$ has been performed considering a fast detector simulation with Delphes at the 3 TeV center-of-mass energy of CLIC with an integrated luminosity of 5.0 ab$^{-1}$ and the 95\% C.L. limits were reported as [-0.03; 0.03] and [-0.007; 0.007] for $\tilde{c}_{HB}$ and $\tilde{c}_{HW}$, respectively \cite{Karadeniz:2020yvz}. Comparing our results with the limits of Ref.~\cite{Karadeniz:2020yvz}, we obtain 1.5 and 2.7 times better sensitivity for CP-violating parameters $\tilde{c}_{HB}$ and $\tilde{c}_{HW}$, respectively. Our results, combined with muon collider's clean experimental environment, the high center-of-mass energy and integrated luminosities, make it a highly promising platform to probe new physics beyond the SM and this is the main motivation for designing this study at future muon collider.

In conclusion, our study demonstrates that future muon colliders, operating at $\sqrt{s}=10$ TeV with high integrated luminosity, will play a crucial role in advancing our understanding of the Higgs sector, particularly in scenarios involving CP-violating Higgs-gauge boson interactions. The enhanced sensitivity to $\tilde{c}_{HB}$ and $\tilde{c}_{HW}$ couplings, along with competitive results for $\tilde{c}_{\gamma}$, underscores the potential of such colliders to explore physics beyond the SM. Future experiments at lepton colliders, combined with complementary results from hadron colliders, will provide a comprehensive approach to probing the fundamental nature of Higgs-gauge boson interactions and their possible CP-violating extensions.

\section{Data Availability Statement}

This manuscript has no associated data or the data will not be deposited. [Authors’ comment: Data will be made available upon reasonable request.]

\end{document}